# WSe$_2$ light-emitting tunneling transistors with enhanced brightness at room temperature

*F. Withers* [1,2]†, *O. Del Pozo-Zamudio*[3]†, *S. Schwarz*[3], *S. Dufferwiel*[3], *P. M. Walker*[3], *T. Godde*[3], *A. P. Rooney*[4], *A. Gholinia*[4], *C. R. Woods*[1], *P. Blake*[1,2], *S. J. Haigh*[4], *K. Watanabe*[5], *T. Taniguchi*[5], *I. L. Aleiner*[6,2], *A. K. Geim*[1], *V. I. Fal'ko*[1,2], *A. I. Tartakovskii*[3], *K. S. Novoselov*[1,2]*

[1]School of Physics and Astronomy, University of Manchester, Oxford Road, Manchester, M13 9PL, UK

[2]National Graphene Institute, University of Manchester, Oxford Road, Manchester, M13 9PL, UK

3 School of Physics and Astronomy, University of Sheffield, Sheffield S3 7RH, UK

[4]School of Materials, University of Manchester, Oxford Road, Manchester, M13 9PL, UK

[5]National Institute for Materials Science, 1-1 Namiki, Tsukuba 305-0044, Japan

[6]Physics Department, Columbia University, New York, New York 10027, USA

†Authors contributed equally

## Abstract

Monolayers of molybdenum and tungsten dichalcogenides are direct bandgap semiconductors, which makes them promising for opto-electronic applications. In particular, van der Waals heterostructures consisting of monolayers of MoS$_2$ sandwiched between atomically thin hexagonal boron nitride (hBN) and graphene electrodes allows one to obtain light emitting quantum wells (LEQW's) with low-temperature external quantum efficiency (EQE) of 1%. However, the EQE of MoS$_2$ and MoSe$_2$-based LEQW's shows behavior common for many other materials: it decreases fast from cryogenic conditions to room temperature, undermining their practical applications. Here we compare MoSe$_2$ and WSe$_2$ LEQW's. We show that the EQE of WSe$_2$ devices grows with temperature, with room temperature EQE reaching 5%, which is 250x more than the previous best performance of MoS$_2$ and MoSe$_2$ quantum wells in ambient conditions. We attribute such a different temperature dependences to the inverted sign of spin-orbit splitting of conduction band states in tungsten and molybdenum dichalcogenides, which makes the lowest-energy exciton in WSe$_2$ dark.

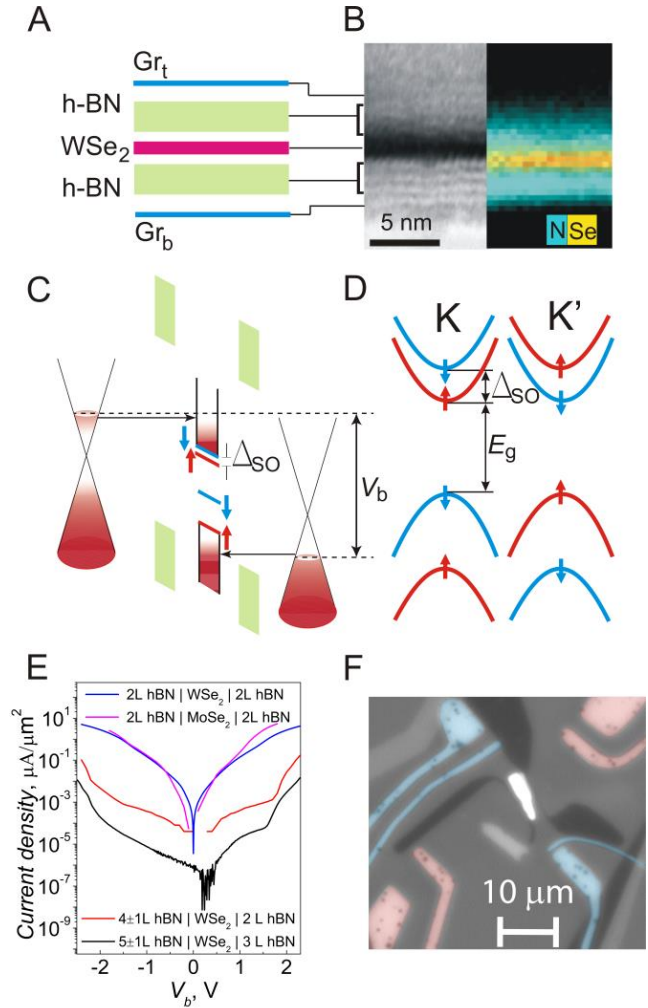

**Figure 1. (A)** Schematic of the device architecture. **(B)** High resolution transmission electron microscopy image of a cross-sectional slice of a WSe$_2$ LEQW on a DBR substrate. **(C)** Band alignment at high bias of a WSe$_2$ LEQW. **(D)** Schematic representation of the band structure of WSe$_2$. Red and blue arrows denote spin orientation. **(E)** Current density vs bias voltage $V_b$ for the presented devices. **(F)** 50x magnification monochrome image of a WSe$_2$ LEQW device with an applied bias of $V_b$=2 V and current of 2 µA taken in ambient conditions with weak backlight illumination. Red false color: Au contacts to bottom graphene, Blue false color: Au contacts to top graphene (Central white region corresponds to strong electroluminescence). See supplementary information for fabrication details.

## Introduction

Recently molybdenum and tungsten dichalcogenides[1] (referred below as MoX$_2$ and WX$_2$ (X= S,Se), respectively) have attracted considerable attention following the discovery of the indirect-to-direct bandgap transition[1-4] and the coupling of the spin and valley degrees of freedom in atomically thin layers[5, 6]. Both in WX$_2$ and MoX$_2$ electrons and holes form strongly bound excitons[7-9] which are stable at room temperature[2-4, 10, 11]. Such properties are very attractive for optoelectronic applications including photovoltaics[12-15] and light emitting diodes[10, 16-23] as well as optical micro-cavity devices[24, 25] including lasers and exciton-polariton structures[26]. Furthermore, a strong spin-orbit interaction in these compounds, has been predicted by density functional theory[27, 28]: in WX$_2$ (MoX$_2$) the lowest energy states in the conduction band and the highest energy states in the valence band have the opposite (same) spin orientations, preventing (enabling) their recombination with emission of a photon (see Fig.1F). Thus, according to the theoretically predicted spin-state ordering, the lowest-energy excitonic sub-band in WX$_2$ corresponds to dark excitons, separated from the bright exciton sub-band by the energy combined from the electron spin-orbit splitting $\Delta_{SO}$ (of the order of 30-40 meV[27-30]) and electron-hole exchange interaction energy. As we show here, such band-structure properties of $WX_2$ strongly impact on the LEQW performance, leading to a significant enhancement of the room temperature EQE of the $WSe_2$ LEQW's in the electroluminescence (EL) regime. This is in contrast to a more common behavior observed in $MoX_2$ LEQW's[10] where the EL intensity falls by up to 100 times when the temperature is varied from 6 to 300 K leading to significant reduction of the EQE.

## Experimental procedure

The main component of our stacked-layer van der Waals heterostructure LEQWs is a light-emitting monolayer of $WSe_2$ encapsulated between thin (2-5 monolayers) hexagonal boron nitride (hBN) barriers with top and bottom transparent graphene electrodes for vertical current injection (Fig. 1). The layer stacks in the van der Waals structure were manufactured using multiple 'peel/lift' procedure[31, 32] in ambient conditions. The high quality of the samples is confirmed by cross-sectional TEM measurements (Fig. 1B), which demonstrate micron scale absence of contamination between the layers occurring as a result of the self-cleansing effect[32, 33] (see supplementary information for AFM and dark field optical microscopy of different devices). We also fabricate similar LEQW structures comprising $MoSe_2$ monolayers. Optical properties of the LEQW devices are studied using micro-photoluminescence (µPL) at low bias voltages, typically $V_b<1.8$ V (or micro-electroluminescence (µEL) measured at larger biases, typically $V_b>1.8$ V) with samples placed in a variable temperature cryostat (see Methods).

By applying a bias voltage, $V_b$, between the two graphene electrodes we are able to pass a tunnel current through the device (Fig. 1E), with the magnitude of the current determined by the largest thickness of one of the hBN barriers. Fig.1E shows the current density (J) vs bias voltage ($V_b$) for four devices having different hBN barrier thicknesses. At high bias we are able to simultaneously inject electrons and holes into the transition metal dichalcogenide (TMDC) layer, Fig. 1C, which is observed as a strong increase of the tunnel conductivity. The lifetime of the injected carriers within the active region of the quantum well is expected to vary as $\tau \propto \theta^{-N}$ where $\theta$ is the probability of an electron tunneling a single layer of hBN and N is the number of hBN atomic layers[34-36] (denoted L below). If the lifetime is long enough then excitons can form within the TMDC and recombine radiatively (Fig.1F). For hBN thickness below 2 L a high proportion of current will be created by direct carrier tunneling through the whole heterostructure leading to a reduction of the current-to-light conversion efficiency. For thicker barriers the lifetime of the carriers increases, leading to improved light emission efficiency. However, in this case the maximum current density drops leading to dimmer LEQW's. We find that 2-3 layers of hBN is an optimal compromise between overall brightness and efficiency ($eN_{ph}/I$) of our devices.

## Results

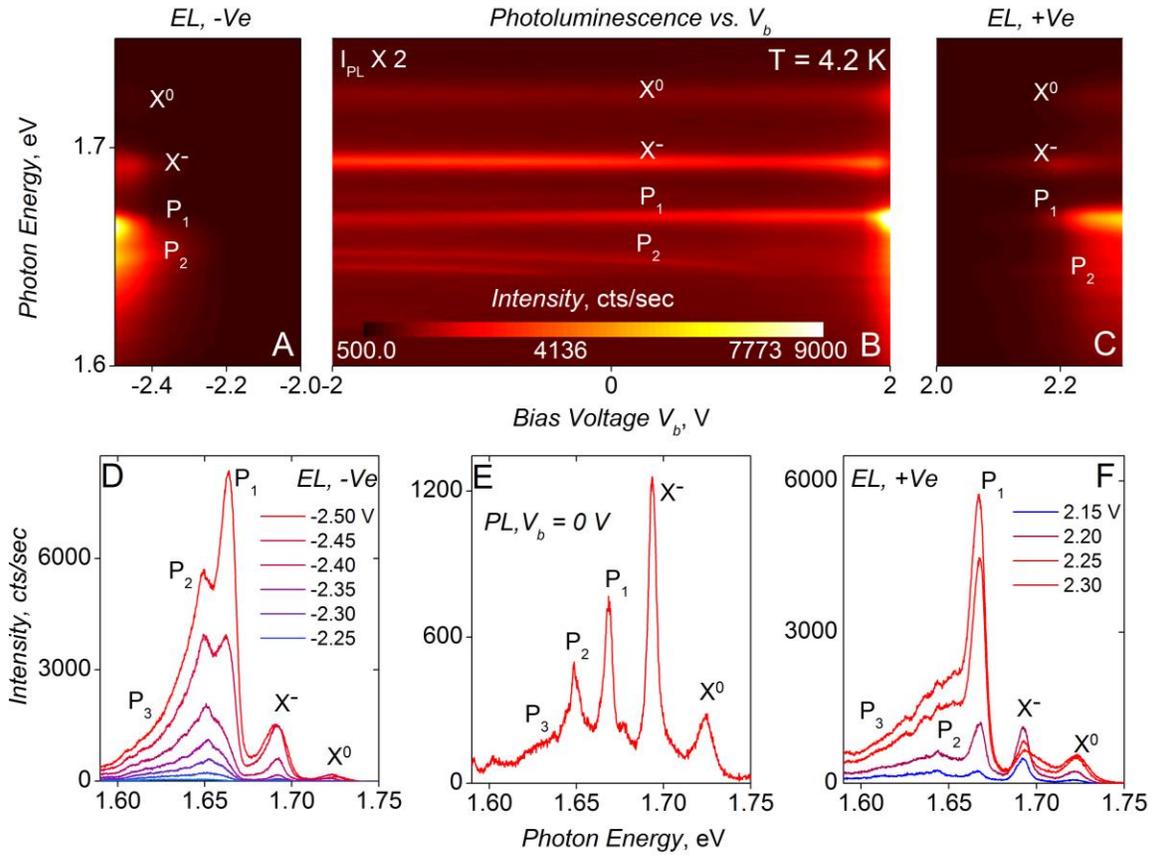

**Figure 2**. Contour maps of the EL spectra from a 5+/- 1L hBN- WSe$_2$-3L hBN LED at T= 4.2 K for negative **(A)** and positive **(C)** bias voltage. **(B)** PL contour map as a function of $V_b$, measured at an excitation power of 10 µW and excitation energy of 1.95 eV. Current density vs bias voltage for this device is shown in Fig. 1E (black curve) **(E)** PL spectrum at $V_b$=0V. **(D), (F)** Bias dependence of the EL for negative and positive polarities respectively (Low bias spectra are presented more clearly in the supplementary materials).

Typical light-emission behavior of a WSe$_2$ LEQW at T = 4.2 K is shown in Fig. 2. PL bias-dependence is shown in Fig. 2B where at zero bias we measure a spectrum shown in Fig. 2E exhibiting several peaks. We use notation adopted in You et al[37], and similarly observe neutral exciton X$^0$ (1.725eV) and trion X$^-$ (1.69 eV) peaks as well as a number of additional features, with the most pronounced peaks P$_1$ (1.669 eV) and P$_2$ (1.649 eV) and a band at photon energies below 1.64 eV denoted P$_3$. We would like to stress that, although features P$_1$-P$_3$ are always present in all our samples, their relative intensity varies from device to device. We attribute the low energy peaks to excitons localized on defects in the TMDC: This is in agreement with theoretical prediction, that their intensity decays faster upon heating than the intensity of the trion line[38]

For biases $|V_b|$>2V, we observe strong electroluminescence (EL). In Figs. 2A,C,D,F the peaks in EL can be easily traced to the peaks in PL spectra, as only insignificant energy shifts of ~3 meV of the spectral features compared with PL occur in the whole range of biases. Figure. 2E shows that at low $T$ with increasing $|V_B|$, EL is first observed from the lowest energy peak $P_3$, then from $P_2$, gradually achieving the maximum through the strong increase of $P_1$ (for clarity we present the low bias behavior in the supplementary information). The high energy EL peaks $X^0$ and $X^-$ also grow with $|V_B|$. Also, at low temperatures, their intensities are somewhat weaker than that of $P_1$.

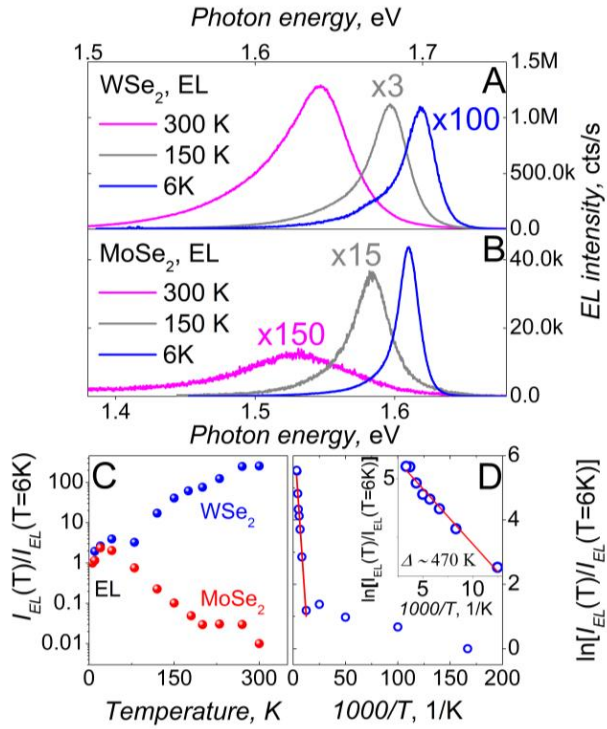

**Figure 3.** **(A)** Electroluminescence spectra taken at different temperatures for a $WSe_2$ QW with 2L hBN tunnel barriers measured with applied bias of $V_b=2V$ (J-$V_b$ is shown in Fig.1E with a blue curve). This sample demonstrates a 200 times increase of the EL output when the temperature is increased from T= 6 K to 300K. **(B)** Electroluminescence spectra recorded for $V_b = 1.8V$ for various temperatures for a $MoSe_2$ LEQW (J-$V_{b\ is}$ shown in Fig.1E with a magenta curve) having identical structure to the $WSe_2$ device in **(A)**. The device shows a typical decrease of light emission with increased temperature. **(C)** Temperature dependence of the integrated EL intensity for a $WSe_2$ (blue) and $MoSe_2$ (red) showing opposite trends with increasing T. The intensities are normalized by those measured for each device at T=6K. **(D)** Arrhenius plot of the EL yield with temperature for the $WSe_2$ device shown in **(A)**. Inset shows the high temperature region used for the linear fit.

It is evident from Fig. 2B that this device shows an asymmetric behavior for positive and negative biases, especially notable in EL (Figs. 2A,C,E,F) as well in the tunnel current (Fig. 1E, black curve). This is typical for the majority of van der Waals LEQWs, likely, due to the different thickness of the top and bottom hBN barriers. EL spectra from the symmetric $WSe_2$ LEQW device are presented in the supplementary information and show similar EL and bias dependent tunnel current for positive and negative polarities at low temperature.

Considering that the key for applications is the room temperature operation of the device, in Fig. 3 we show temperature dependence of the EL comparing two LEQWs, one with monolayer of $MoSe_2$ and the other with monolayer $WSe_2$, with the same architecture. The dependence of the tunnel current density on bias voltage for these two devices, Fig. 1E, shows both samples that the current density at a given bias voltage is of the same order of magnitude.

The emission of the WSe$_2$-based device becomes around 200 times brighter at 300K, as compared to T=6 K Fig. 3A. Qualitatively similar behavior has been observed in the previous PL studies on monolayers of WSe$_2$: typically a factor of ten[39, 40] increase in the range from 6 K to 300 K. Such PL variation has also been observed for the WSe$_2$ device shown in Fig. 3A (see SI). We have checked that such unusual temperature dependence of PL and EL is reproduced in a multiplicity of WSe$_2$ LEQW devices. Depending on a particular device, the integrated EL increase from T = 6 to 300 K is in the range from 4 to 200 times, and the increase of 200 times occurs in the samples with thin 2-3 L hBN barriers. In such devices the trion emission dominates both in PL and EL and X$^0$ emission is not observed (Fig. 3A), leading to a simpler spectrum with one peak at 1.698 eV and a shoulder at 1.666 eV for T=6 K. Note that such samples also provide the brightest EL exceeding 1 million counts per second at room temperature (see Methods for details of the LEQW substrate leading to enhanced collection efficiency).

In Fig. 3(B-D) we also carry out direct comparisons with a MoSe$_2$ LEQW fabricated in an identical way to the WSe$_2$ devices discussed above. As shown in Fig. 3B,C, the brightest EL from MoSe$_2$ devices is observed at low *T*, where a single EL peak is observed close to the spectral position of trion emission peak (see Supplementary Information for bias dependent PL characterization). As the temperature is increased, this peak significantly broadens, and the integrated EL intensity decreases by about 100 times, see Fig. 3C. This is similar to the earlier found ten-fold decrease in the EL intensity of MoS$_2$ LEQWs between 10 K and 300 K[10].

The strong increase of EL with temperature corresponds to a rise of the external quantum efficiency (EQE) in WSe$_2$ LEQWs and follows an Arrhenius relation with an energy gap of 40 meV (see Fig. 3D). This makes van der Waals heterostructures with embedded WSe$_2$ monolayers highly promising material for ultra-thin flexible LEQWs. Figure.4 shows the EQE T-dependence for three WSe$_2$ devices (data for additional devices are shown in the SI). Here the EQE is defined as *EQE = eN$_{ph}$/I*, where *N$_{ph}$* is the number of photons emitted by the device, *I* is the current passing through the device and *e* is the electron charge. It is observed in Fig. 4A that the EQE shows the characteristic increase with temperature reaching 5% at T = 300 K, a factor 250 improvement in the room temperature performance as compared to the best single-monolayer MoS$_2$ LEQWs[10].

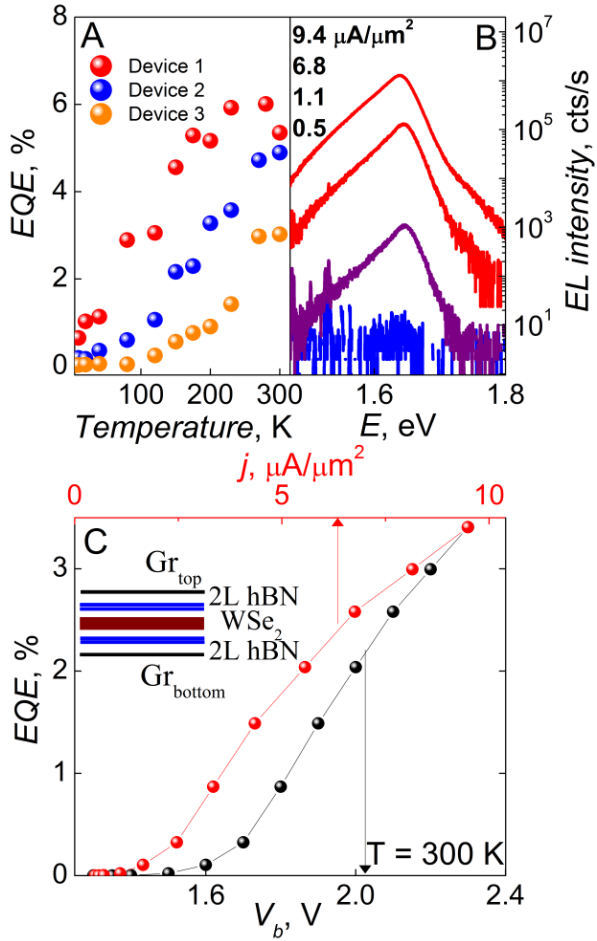

In addition to this, in Fig. 4C we observe a monotonically increasing EQE as a function of bias voltage and injection current density to a maximum measured value of $j=10^3$ A/cm$^2$. Fig. 4B shows the corresponding EL spectra obtained from Device 3 in Fig. 4A operated at room temperature at increasing injection current densities with a peak emission of more than 1.3 million counts per second. Indeed, a common drawback of commercial LED lighting is a suppression of the EQE (so-called 'efficiency droop') at high injection currents caused by increased non-radiative processes[41, 42]. In contrast, the presented WSe$_2$-based van der Waals heterostructure LEQWs become brighter at higher temperature, and their efficiency remains high and increases with current densities even at current densities as high as 1000 A/cm$^2$ as shown in Fig. 4C.

We suggest that the mechanism of the unusual $T$-dependence of EL and PL in WSe$_2$ LEQWs is related to the spin-orbit splitting of the spin states in the free carrier bands as illustrated in Fig. 1D, leading to the specific sequence of bright and dark exciton states in these materials[27, 28]. In the discussion below we will focus on the behavior of the neutral exciton which dominates in PL and EL at room T in the majority of our LEQWs. The dark and bright exciton states are split by $\Delta = \Delta_{SO} + \Delta_{e-h}$, where $\Delta_{SO} \approx 30\text{-}40$ meV[27-30] is the spin-orbit splitting of electrons in the conduction band (which has the opposite sign to that of MoX$_2$) and $\Delta_{e-h} \approx 1.5$ meV is the interband (conduction-valence) exchange energy.

**Figure 4.** **(A)** Temperature dependence of the quantum efficiency for three typical WSe2 LED devices measured at bias voltages and injection currents of 2.8 V and j = 0.15 μA/μm2 (Device 1), 2.8 V and j = 0.5 μA/μm2 (Device 2), 2.3 V and j = 8.8 μA/μm2 (Device 3). **(B)** Individual electroluminescence spectra plotted for four different injection current densities for Device 3. **(C)** The external quantum efficiency plotted against bias voltage and injection current density at T = 300 K for Device 3. The EQE monotonically increases even up to current densities of 10 μA/μm2 or 1000 A/cm2.

At low temperature the exciton population accumulates in the low energy dark exciton sub-band, which mostly decays via non-radiative escape. At the same time, at low T the bright exciton states are weakly populated as it is shifted to higher energy by Δ with respect to the dark exciton, hence, the intensity of the $X^0$ line is low. As the temperature increases, the thermal activation increases the bright exciton population. This population will be mostly contained in the high-momentum exciton 'reservoir', whereas light emission occurs from exciton states with the negligible momentum ($k \approx 0$). Excitons from the 'reservoir' can be scattered into the light-emitting states as a result of interaction with acoustic phonons and electrons, (the latter is particularly significant in the EL regime), or defects.

The fact that, at high *T,* the intensity of the $X^0$ line increases both in PL and EL - is the manifestation of such thermal activation behavior. Note that for several of the studied devices the Arrhenius fit to the exponential increase of the EL with increasing temperature yields the characteristic energy of ~40meV (inset to Fig.3D), which is quite close to the theoretically predicted $\Delta_{SO}$[27].

This is in contrast to $MoSe_2$ LEQWs, where the dark exciton is higher in energy than the bright exciton. As a result, $MoSe_2$ LEQW's always show the opposite behavior with a notable decrease of light emission with increasing *T*. While it is expected that the emission efficiency would decrease due to the transfer of some exciton population into the dark exciton sub-band and also into the 'reservoir' states, such a strong decrease by a factor of 100 and more may only be possible if the non-radiative escape time shortens at elevated temperature, which appears to be a stronger effect in $MoSe_2$ compared with $WSe_2$.

In conclusion, we have fabricated high-efficiency LEQWs made from van der Waals heterostructures comprising a single layer of $WSe_2$ as the active light-emitting material, hBN tunnel barriers, and graphene electrodes for vertical current injection. Such $WSe_2$-based LEQWs show the enhanced performance at room temperature compared with the low-*T* operation. This enhancement is also in contrast to $MoSe_2$ and $MoS_2$ LEDs studied in this work and reported earlier, where both PL and EL decrease by a factor of 10 to 100 when the temperature is varied from 6 to 300 K. With room temperature external efficiencies of 5%, such LEDs present significant promise for future development of flexible opto-electronic components. The efficiency can be boosted further by creating 'multiple quantum well' devices[10] and by the fine tuning of the h-BN tunnel barrier thickness. One of the remaining challenges is scalable production of these components, only possible with well-controlled wafer-scale growth techniques[43, 44].

**Materials and Methods**

LED fabrication: Firstly bulk hexagonal boron nitride hBN is mechanically cleaved and exfoliated onto a freshly cleaned Si/SiO2 substrate. After this a graphene flake is peeled from a PMMA membrane onto the hBN crystal followed by a thin hBN tunnel barrier then a hBN tunnel barrier on PMMA is used to lift a WSe2 or MoSe2 single layer from a second substrate then both of these crystals are together peeled off the PMMA onto the hBN/Gr/hBN stack forming hBN/Gr/hBN/WX2/hBN. Finally the top graphene electrode is peeled onto the stack thus completing the LED structure (see supplementary information for fabrication details). After the stack is completed we follow standard micro fabrication procedures for adding electrical contacts to the top and bottom graphene electrodes. We also transferred some complete stacks onto highly reflective distributed Bragg reflector (DBR) substrates. From LEDs placed on such DBRs we are able to collect up to 30% of the emitted light opposed to just 2% from the Si/SiO2 substrate.

Electrical and optical measurements: Samples were mounted either in a liquid helium flow cryostat for temperature dependent measurements, or in an exchange-gas cryostat for measurements at T= 4.2 K. Light emission from the samples was collected with a high numerical aperture lenses positioned either outside the flow cryostat or inside the exchange-gas cryostat. The photoluminescence (PL) and electroluminescence (EL) signals were measured using a 0.5m spectrometer and a nitrogen cooled charge coupled device (Princeton Instruments, Pylon CCD). Electrical injection is performed using a Keithley 2400 source-meter. PL was excited with continuous wave lasers at 532 nm or 637 nm focused in a spot size of ~ 2 to 3 μm on the sample surface. The optical image taken in Figure 1F was captured using a Nikon DS-Qi2 monochrome camera with a quantum efficiency of 20% at 750 nm.

**Scanning transmission electron microscopy (STEM):** STEM imaging was carried out using a Titan G2 probe-side aberration-corrected STEM operating at 200 kV and equipped with a high-efficiency ChemiSTEM energy-dispersive X-ray detector. The convergence angle was 19 mrad and the third-order spherical aberration was set to zero (±5 μm). The multilayer structures were oriented along the <hkl0> crystallographic direction by taking advantage of the Kikuchi bands of the silicon substrate. (See Supplementary information and ref.19 for more detailed description).

ASSOCIATED CONTENT

Supporting Information: Description of the device fabrication, Temperature dependent electroluminescence data for additional WSe2 LED's, low temperature photoluminescence of MoSe2 LED's, details of our quantum efficiency estimations and further discussion. "This material is available free of charge via the Internet at http://pubs.acs.org."

AUTHOR INFORMATION

**Corresponding Author**
*Correspondence and requests for materials should be addressed to K.S.N. (kostya@manchester.ac.uk), A.I.T. (a.tartakovskii@sheffield.ac.uk) and F.W. (freddie.withers@manchester.ac.uk).

**Author Contributions**
F.W. produced experimental devices, measured device characteristics, analyzed experimental data, participated in discussions, contributed to writing the manuscript; O.D.P.Z., S.S., S.D. and T.G. measured device characteristics, analyzed experimental data, participated in discussions; P.M.W. calculated light collection efficiency, analyzed experimental data, participated in discussions; A.P.R. and A.G. produced samples for TEM study, analyzed TEM results, participated in discussions; K.W. and T.T. grew high quality hBN, participated in discussions; S.J.H. produced samples for TEM study, analyzed TEM results, participated in discussions; A.K.G. analyzed experimental data, participated in discussions, contributed to writing the manuscript; A.I.T. measured device characteristics, analyzed experimental data, participated in discussions, contributed to writing the manuscript; V.F. has proposed the interpretation of the spectra and contributed to writing the manuscript; K.S.N. initiated the project, measured device characteristics, analyzed experimental data, participated in discussions, contributed to writing the manuscript.

**Notes**

The authors declare no competing financial interest.

ACKNOWLEDGMENT

This work was supported by European Research Council Synergy Grant Hetero2D, EC-FET European Graphene Flagship, The Royal Society, Royal Academy of Engineering, U.S. Army, European Science Foundation (ESF) under the EUROCORES Programme EuroGRAPHENE (GOSPEL), Engineering and Physical Sciences Research Council (UK), the Leverhulme Trust (UK), U.S. Office of Naval Research, U.S. Defence Threat Reduction Agency, U.S. Air Force Office of Scientific Research, FP7 ITN S3NANO, SEP-Mexico and CONACYT.

# WSe$_2$ light-emitting tunnelling transistors with enhanced brightness at room temperature

F. Withers[1]*, O. Del Pozo-Zamudio[2]*, S. Schwarz[2], S. Dufferwiel[2], P. M. Walker[2], T. Godde[2], A. P. Rooney[3], A. Gholinia[3], C. R. Woods[1], P. Blake[1,5], S. J. Haigh[3], K. Watanabe[4], T. Taniguchi[4], I. L. Aleiner[5,6,7], A. K. Geim[8], V. I. Fal'ko[1,6,7], A. I. Tartakovskii[2], K. S. Novoselov[1]

[1]School of Physics and Astronomy, University of Manchester, Oxford Road, Manchester, M13 9PL, UK

[2]Department of Physics and Astronomy, University of Sheffield, Sheffield S3 7RH, UK

[3]School of Materials, University of Manchester, Oxford Road, Manchester, M13 9PL, UK

[4]National Institute for Materials Science, 1-1 Namiki, Tsukuba 305-0044, Japan

[5]Manchester Centre for Mesoscience and Nanotechnology, University of Manchester, Oxford Road, Manchester, M13 9PL, UK

[6]National Graphene Institute, University of Manchester, Oxford Road, Manchester, M13 9PL, UK

[7]Department of Physics, Lancaster University, Lancaster, LA1 4YB, UK

*Authors contributed equally

# Fabrication

Quantum well heterostructure devices are assembled by a multiple peel-lift Van der Waals assembly procedure which has been described in detail previously [1-3].

In summary the devices are constructed as follows, firstly an hBN flake of thickness 5-35 nm is deposited onto a thermally oxidized silicon wafer (90 or 290 nm oxide thickness) to form an atomically flat substrate. A graphene flake is then peeled from a poly(methyl methacrylate) (PMMA) membrane onto the hBN substrate, followed by a thin hBN tunnel barrier of thickness 2-5 monolayers (L).

Another hBN tunnel barrier of thickness 2-5 L on a PMMA membrane is then used to lift (by Van der Waals forces) a single layer of transition metal dichalcogenide (TMDC) from a separate $SiO_2$ substrate. The hBN together with the single layer TMDC is then peeled onto the already constructed hBN-Gr-hBN structure to form the stack of hBN-Gr-hBN(2-5L)-TMDC(1L)-hBN(2-5L). Finally a graphene flake is then peeled from a membrane to complete the light-emitting quantum well (LEQW) device. Estimation of the hBN tunnel barrier thickness is conducted using a combination of optical and atomic force microscopy measurements.

Electrical contacts are patterned using electron beam lithography followed by evaporation of Cr/Au (5nm/50nm) allowing for independent electrical contacts to both the top and bottom graphene electrodes.

Devices were also fabricated onto distributed Bragg reflector (DBR) substrates which allow for the collection of 30% of the emitted light and leads to much brighter LEQW's. Details of the Bragg reflector substrates can be found in [4, 5]. Heterostructure LEQW's on DBR's were firstly fabricated on a thermally oxidized Si wafer then the whole heterostructure stack was transferred by using the wet transfer method from the $SiO_2$ substrate to the DBR mirror followed by e-beam lithography and metallization. This step was found to be necessary due to poor adhesion of flakes to the DBR mirrors preventing direct exfoliation onto the mirror surfaces. The DBR dielectric pairs were also found to delaminate when removing tape during mechanical exfoliation.

All TMDC materials were sourced from either **HQ-Graphene** or **2Dsemiconductor**.

1. Fabrication steps for the glowing device shown in Figure 1 of the main text

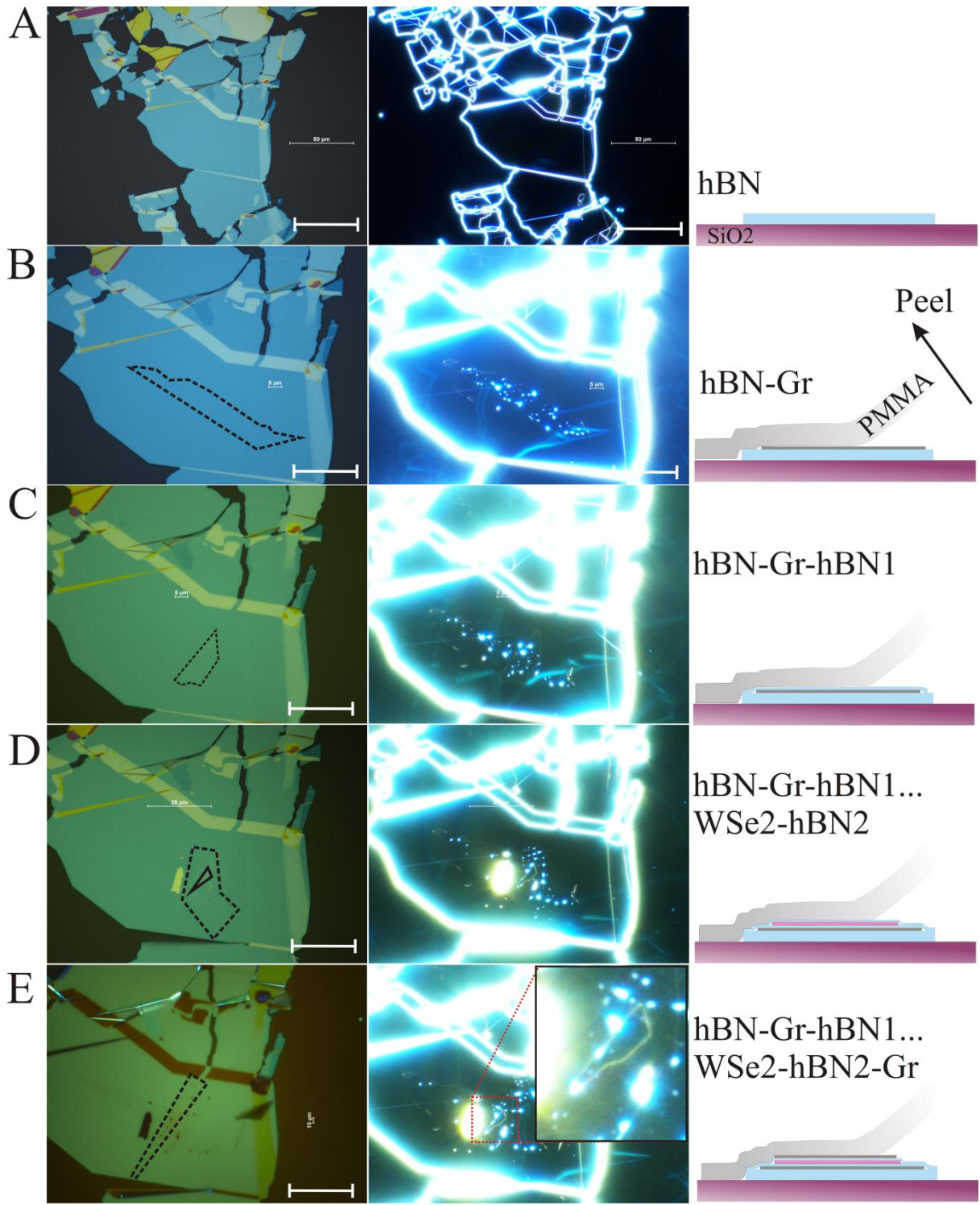

Figure S1. **A**, hBN crystal exfoliated onto an oxidized silicon wafer, dark field images are shown on the right; **B**, a graphene flake is peeled from an PMMA membrane onto the large hBN crystal; **C**, the graphene flake is then covered with the first hBN tunnel barrier which is again peeled from a PMMA membrane. **D**, the quantum well is completed by using a second hBN tunnel barrier to lift a $WSe_2$ flake from a separate Si-SiO$_2$ substrate, the second tunnel barrier together with the $WSe_2$ layer are then peeled onto the hBN-Gr stack. **E**, Finally the top graphene electrode is then peeled completing the heterostructure stack (inset: shows a blown up image of the heterostructure region). Scale bars 50 μm **in A,** 25 μm **in B-E.**

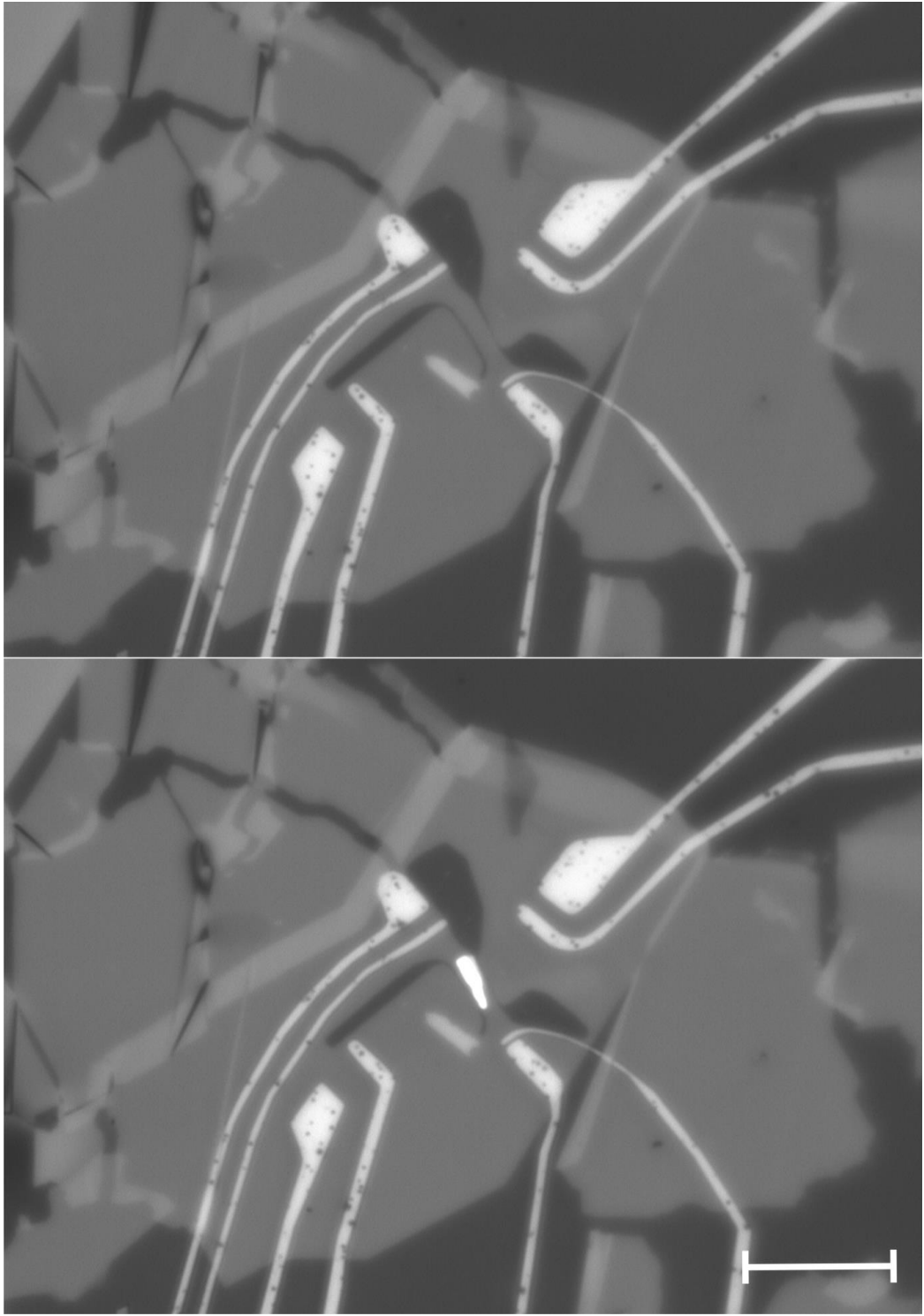

*Figure S2.**Top**: Optical micrograph of the of the completed device, the dark regions either side of the overlap region correspond to an etch mesa in PMMA used to remove excess graphene outside the overlap region. **Bottom**: image of the device under a bias of $V_b$ = 2 V. The central overlap region shows strong electroluminescence. (Scale bar: 25 μm)*

2. Atomic force microscopy showing self-cleaning mechanism

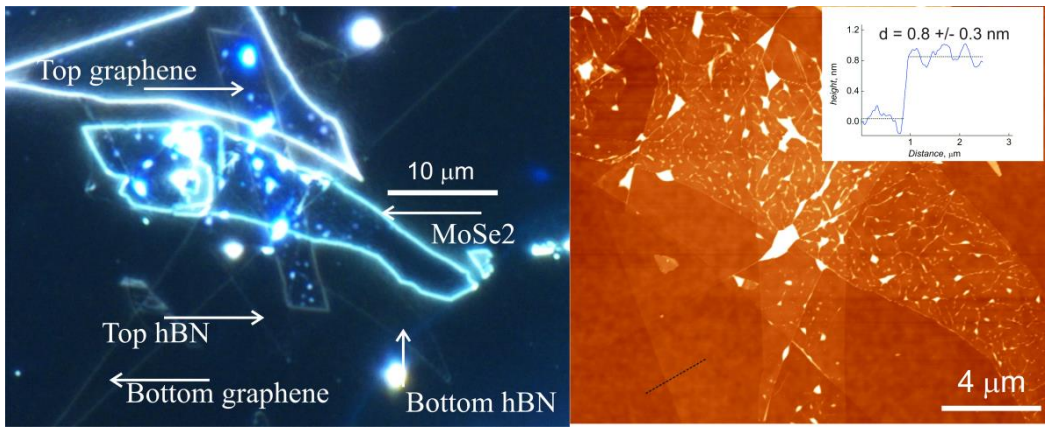

*Figure S3. **Left:** Dark field optical micrograph of a completed quantum well stack with significant contamination introduced during the hBN/MoSe$_2$ peel step. The bright lines correspond to the edges of the flakes while the white spots correspond to pockets of trapped contamination. **Right:** Atomic force microscopy reveals that the trapped contamination self-cleans into pockets leaving ~ µm sized atomically flat regions. **Inset**: AFM step profile used to estimate the number of layers in one of the hBN tunnel barriers.*

## Additional devices and data

1. Temperature dependence of additional WSe$_2$ LED's

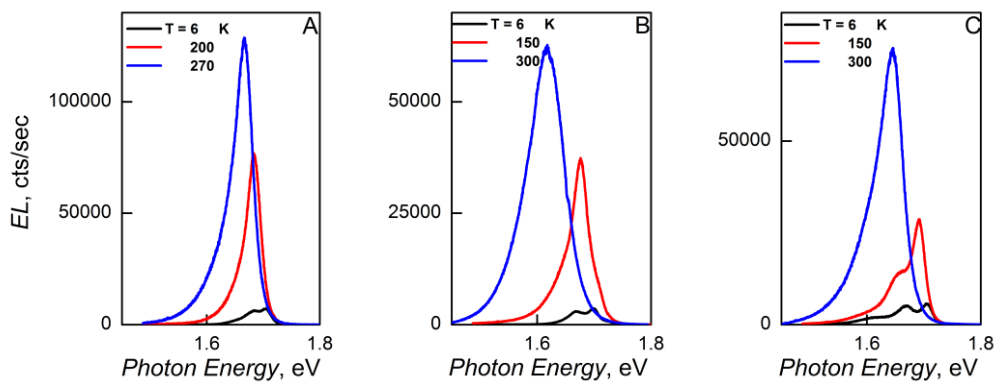

*Figure S4. **A,B,C** Electroluminescence (EL) spectra taken at increasing temperatures for three additional WSe$_2$ LEQW's devices, all of which show the characteristic increase of the electroluminescence output with increasing temperature.*

## 2. Temperature dependence of the Tunneling conductivity

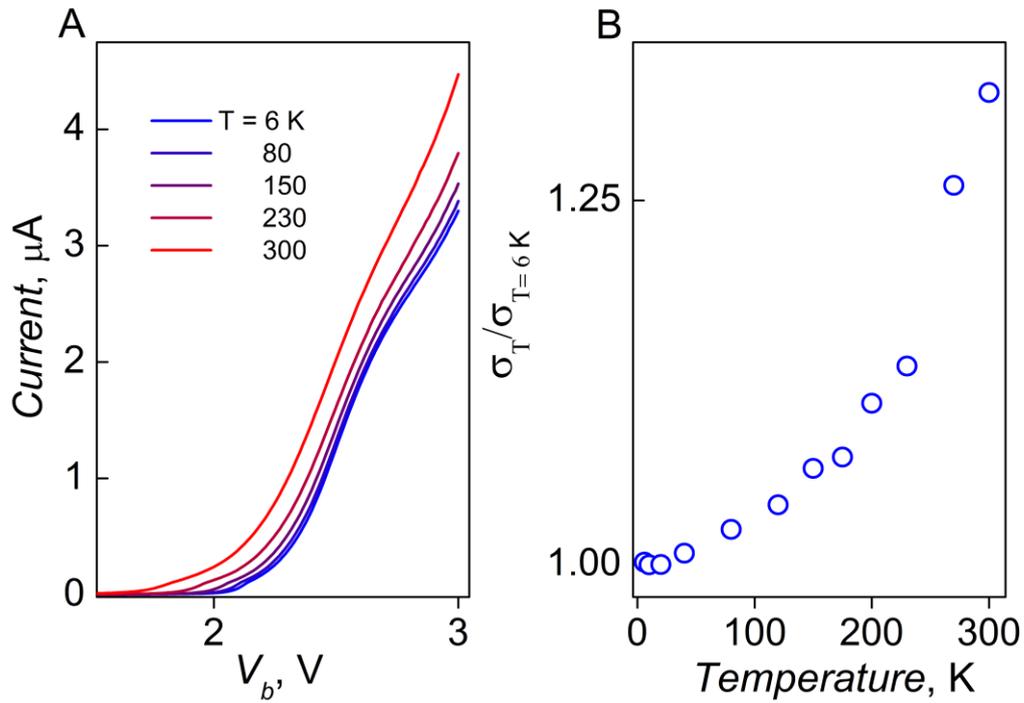

*Figure S5. **A**, Typical I-$V_b$ dependence for a LEQW device, showing only weak dependence on temperature **B**, ratio of the tunnel conductivity at a given temperature to that of T = 6 K showing only a small increase from T = 6 K to T = 300 K taken at $V_b$ = 2.8 V.*

All our LEQW devices display only a weak dependence of the tunnel current on increasing temperature, at most the tunnel current increases by a factor of 2 times in some samples. Figure S5 shows the typical temperature dependent behavior of one of our LEQW's showing a nominal ~ 1.3X increase in current as the temperature is increased from T = 6 K to T = 300 K.

These small changes of conductivity at a fixed bias voltage are taken into account by normalizing the EL spectra at each temperature by the conductivity as follows,

$$I_{Norm}^{EL}(T) = \frac{I^{EL}(T)\sigma(T=6K)}{\sigma(T)}.$$

## 3. Low temperature PL / EL of WSe$_2$ and MoSe$_2$ LEQW's

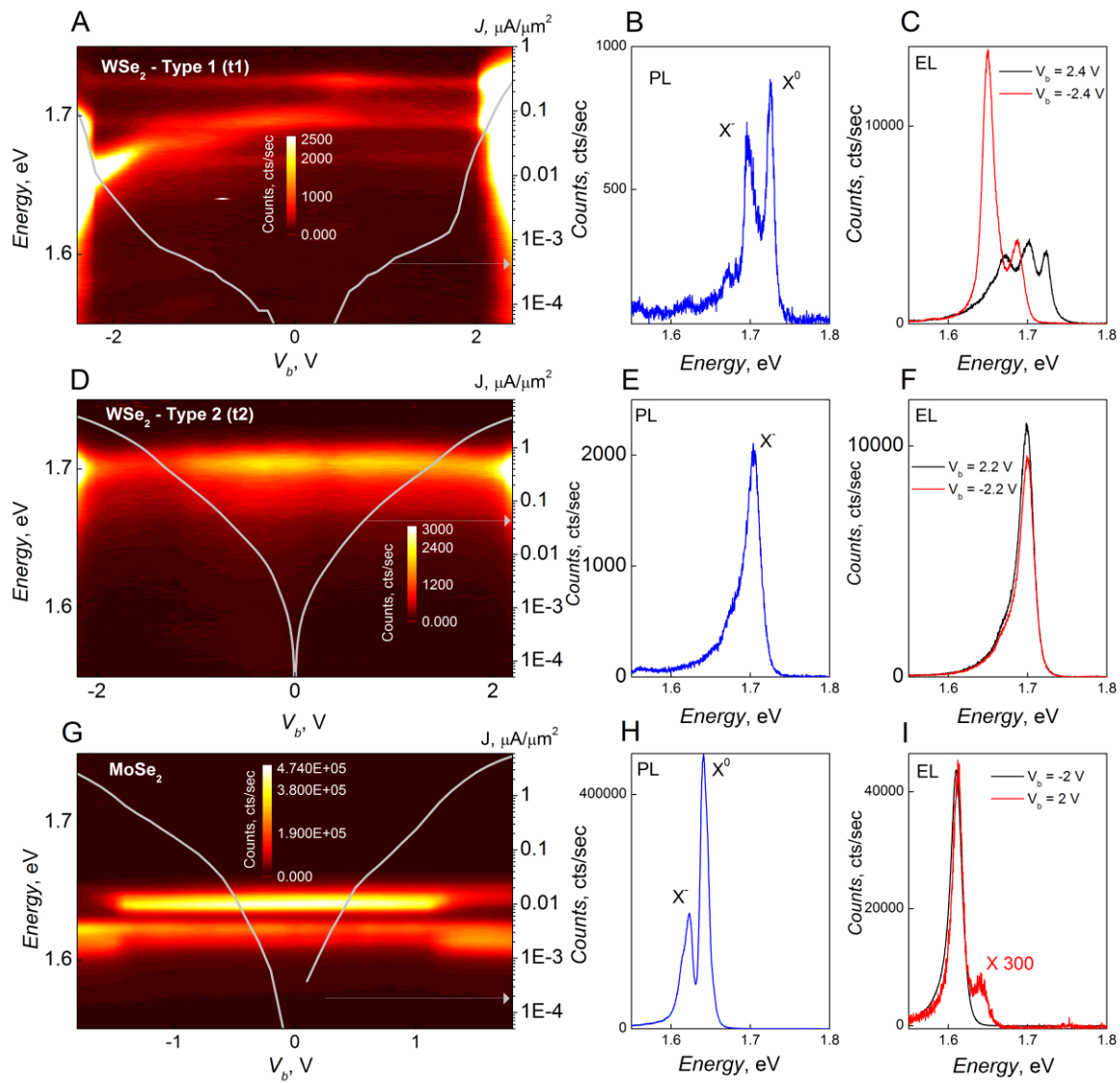

*Figure S6. Photoluminescence (PL) and electroluminescence (EL) spectra measured at T= 6 K. **A**, contour map of PL vs bias voltage for a WSe$_2$ LEQW with the current density shown on the right y-axis and the data plotted in white. **B**, PL spectrum at V$_b$ = 0 V and **C**, EL spectra for positive and negative biases. **D,E,F,** same as in **A,B,C** but for a symmetric LEQW's with 2L hBN tunnel barriers around the WSe$_2$ layer. **G,H,I,** same as in **A,B,C,** but for a symmetric (2L hBN barriers) MoSe$_2$ LEQW's. PL was measured using a 532nm laser at a power of 32 µW.*

## 4. Low bias EL spectra for the device shown in Figure 2 of the main text.

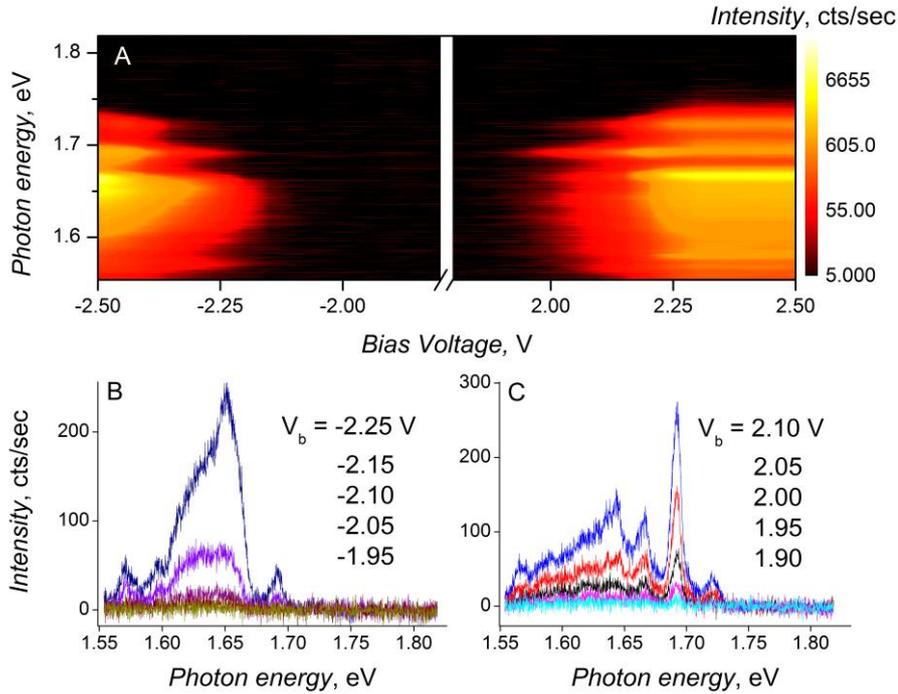

*Figure S7. **A,** Electroluminescence contour maps for positive and negative bias voltage plotted in logarithmic scale to better show the onset of EL. **B,C** Electroluminescence spectra in the low bias regime for positive and negative bias.*

## 5. Temperature dependence of the low energy peaks

Here we consider in more detail the temperature dependence of the low energy peaks $P_1$ and $P_2$ in a typical $WSe_2$ LEQW device. As the temperature increases from 20 to 80 K, the EL intensity of the high energy peak $X^0$ increases by a factor of 2 and the $X^-$ peak grows by 1.5 times (see Fig. S8). At the same time the low energy peaks $P_1$ and $P_2$ start to decay, with $P_2$ showing a sharp decay by a factor of 2. At higher temperatures these low energy peaks start to broaden, merge and it becomes hard to trace individual features. The observed redistribution of the EL intensity clearly shows thermal-activation type behavior where the occupation of the low energy states decreases, while the population of the high energy states grows with temperature. Note, that, eventually, at room *T*, the neutral exciton line dominates in PL and EL in the majority of LEQWs studied in this work (see also PL results on $WSe_2$ monolayer films in Refs.[6, 7]).

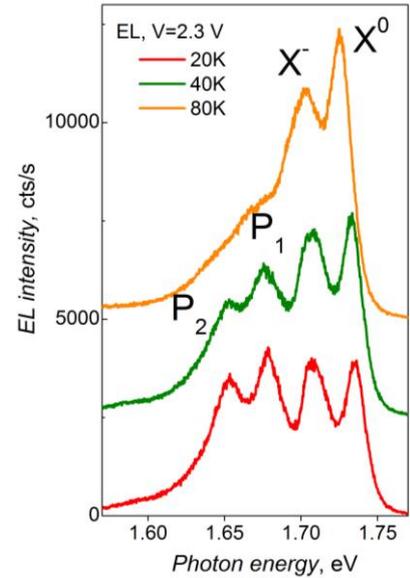

*Figure S8. Comparison of the photoluminescence intensity for $MoSe_2$ and $WSe_2$ devices shown in Figure 3D.*

## 6. Temperature dependence of the photoluminescence for MoSe$_2$ and WSe$_2$

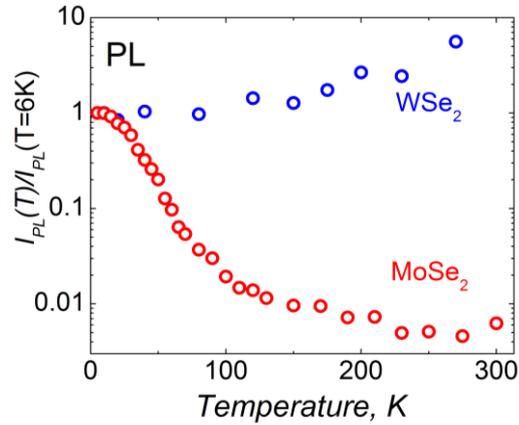

*Figure S9. Comparison of the photoluminescence intensity for MoSe$_2$ and WSe$_2$ devices shown in Figure 3D.*

Figure S9 shows the temperature dependence of the photoluminescence integrated intensity for the MoSe$_2$ (red) and WSe$_2$ (blue) devices shown in figure 3D of the main text.

The PL yield of the MoSe$_2$ device drops off by a similar amount to the EL in this device. However for the WSe$_2$ device the PL intensity only grows by roughly 10x as compared to 200x increase of the EL in this structure.

## 4. Collection efficiency

The quantum efficiency is defined as the number of photons emitted per number of injected carriers, $Ne/i$ ($N$ = number of emitted photons per second, $e$ electron charge, $I$ is the current passing through our collection area). In order to estimate the number of emitted photons we need to estimate our collection efficiency. The total loss is defined as,

$\eta = \eta_{Lens}\eta_{optic}\eta_{system}$.

$\eta_{optic}$ is the loss of all the optical components in the optical circuit. It was measured directly using a 1.96 eV laser and a power meter to determine the loss at each component. We find $\eta_{optic}$ = 0.18.

$\eta_{system}$ - converts the number of photons arriving at the incoming slit of the detector into the detector counts. It takes into account the loss of photons which pass through the slit, grating and onto the CCD and has been again measured directly by using the 1.96 eV laser and taking spectra of the laser for different powers in order to get a counts vs incident photons. For our system we get 4203 integrated cts/sec per 1 pW. Taking into account that 1 pW of power corresponds to N=P/hν=3177476 photons, we arrive at a conversion coefficient between the number of integrated counts and the number of photons incident on the slit of the spectrometer per second leading to the system efficiency of $\eta_{system}$=4203/3177476 =1.32 x 10$^{-3}$.

$\eta_{Lens}$ is the efficiency of the lens collection[8]. We use a 50x objective with a numerical aperture, NA = 0.55. LED's are fabricated on either two substrates, firstly Si-SiO$_2$(290nm) with refractive index of Si(n=3.734) and SiO$_2$(n=1.645) or distributed Bragg reflectors which consist of 10 alternating quarter wave pairs (187.5nm) of SiO$_2$(n=1.46) and NbO$_2$(n=2.122)(see [4, 5]).

Numerical simulations allow us to make an improved estimate of the collected light emitted from a dipole on each substrate type.

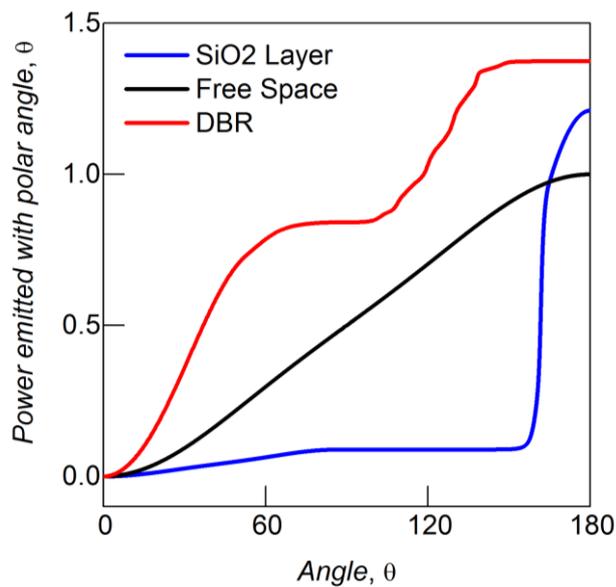

*Figure S10. Total power emitted within a polar angle for an emitting dipole placed on Si-SiO$_2$ and on a distributed Bragg reflector (DBR) as well as in free space.*

Our NA = 0.55 . This gives a collection angle of 33.4 degrees.

So for a WSe$_2$ flake on SiO$_2$ we find that

$$\eta_{Lens} = \frac{P_{SiO2}(33.4)}{P_{SiO2}(180)} = 2.5\ \%$$

While for WSe$_2$ emitting on a DBR we find,

$$\eta_{Lens} = \frac{P_{DBR}(33.4)}{P_{DBR}(180)} = 31.0\ \%$$

This gives us two loss factors depending on the substrate the LED was fabricated onto.

$$\eta_{SiO2} = 0.18 \times 1.32 \times 10^{-3} \times 0.025 = 5.94 \times 10^{-6}$$

$$\eta_{DBR} = 0.18 \times 1.32 \times 10^{-3} \times 0.31 = 7.37 \times 10^{-5}$$

## Cross sectional imaging

Further details of cross sectional imaging of heterostructures produced from 2D materials can be found in [1, 9] .

1. Sample preparation

In summary a dual beam instrument (FEI Dual Beam Nova 600i) has been used for site specific preparation of cross sectional samples suitable for TEM analysis using the lift-out approach [Schaffer, M. *et al.* Sample preparation for atomic-resolution STEM at low voltages by FIB [10]]. This instrument combines a focused ion beam (FIB) and a scanning electron microscope (SEM) column into the same chamber and is also fitted with a gas-injection system to allow local material deposition and material-specific preferential milling to be performed by introducing reactive gases in the vicinity of the electron or ion probe. The electron column delivers the imaging abilities of the SEM and is at the same time less destructive than FIB imaging. For heterostructures fabricated on insulating substrates, such as DBR's, a thin layer of Au was initially deposited to prevent charging when SEM imaging. SEM imaging of the device prior to milling allows one to identify an area suitable for side view imaging. After sputtering of a 10 nm carbon coating and then a 50 nm Au-Pd coating on the whole surface ex-situ, the Au/Cr contacts on graphene were still visible as raised regions in the secondary electron image. These were used to correctly position and deposit a Pt strap layer on the surface at a chosen location, increasing the metallic layer above the device to ~2 µm. The Pt deposition was initially done with the electron beam at 5kV e$^-$ and 1nA up to about 0.5µm in order to reduce beam damage and subsequently with the ion beam at 30kV Ga$^+$ and 100pA to build up the final 2µm thick deposition. The strap protects the region of interest during milling as well as providing mechanical stability to the cross sectional slice after its removal. Trenches were milled around the strap by using a 30 kV Ga$^+$ beam with a current of 1-6nA, which resulted in a slice of about 1µm thick. Before removing the final edge supporting the milled slice and milling beneath it to free from the substrate, one end of the Pt strap slice was welded to a nano-manipulator needle using further Pt deposition. The cross sectional slice with typical dimensions of 1 µm x 5 µm x 10 µm could then be extracted and transferred to an Omniprobe copper half grid as required for TEM. The slice was then welded onto the grid using Pt deposition so that it could be safely separated from the nanomanipulator by FIB milling. The lamella was further thinned to almost electron beam transparency using a 30kV Ga$^+$ beam and 0.1-1nA. A final gentle polish with Ga+ ions (at 5kV and 50pA) was used to remove side damage and reduce the specimen thickness to 20-70nm. The fact that the cross sectional slice was precisely extracted from the chosen spot was confirmed for all devices by comparing the positions of identifiable features such as Au contacts and /or hydrocarbon bubbles, which are visible both in the SEM images of the original device and within TEM images of the prepared cross section.

2. Scanning transmission electron microscope imaging and energy dispersive x-ray spectroscopy analysis

High resolution scanning transmission electron microscope (STEM) imaging was performed using a probe side aberration-corrected FEI Titan G2 80-200 kV with an X-FEG electron source operated at 200kV. High angle annular dark field (HAADF) and bright field (BF) STEM imaging was performed using a probe convergence angle of 26 mrad, a HAADF inner angle of 52 mrad and a probe current of ~200 pA. Energy dispersive x-ray (EDX) spectrum imaging was performed in the Titan using a Super-X four silicon drift EDX detector system with a total collection solid angle of 0.7 srad. The multilayer structures were oriented along an <hkl0> crystallographic direction by taking advantage of the Kukuchi bands of the Si substrate.